\documentclass{emulateapj}

\newcommand{\kms}{km~s$^{-1}$}

\newcommand{\msun}{$M_{\odot}$}

\newcommand{\feh}{[Fe/H]}

\def\f0{$F_0$}

\input epsf

\begin{document}
\title{Carbon-Enhanced Hyper-metal-poor Stars and the Stellar IMF at Low Metallicity}
\author{Jason Tumlinson}
\affil{Yale Center for Astronomy and Astrophysics, Departments
of Physics and Astronomy, Yale University, P. O. Box 208121, New Haven,
CT 06520}
\begin{abstract}
The two known ``hyper-metal-poor'' (HMP) stars, HE0107-5240 and HE1327-2326, have extremely high enhancements of the light elements C, N, and O relative to Fe and appear to represent a statistically significant excess population relative to the halo metallicity distribution extrapolated from [Fe/H] $> -3$. This study weighs the available evidence for and against three hypothetical origins for these stars: (1) that they formed from gas enriched by a primordial ``faint supernova'', (2) that they formed from gas enriched by core-collapse supernovae and C-rich gas ejected in rotation-driven winds from massive stars, and (3) that they formed as the low-mass secondaries in binary systems at $Z \sim 10^{-5.5} Z_{\odot}$ and acquired their light-element enhancements from an intermediate-mass companion as it passed through an AGB phase. The observations interpreted here, especially the depletion of lithium seen in HE1327-2326, favor the binary mass-transfer hypothesis. If HE0107-5240 and HE1327-2326 formed in binary systems, the statistically significant absence of isolated and/or C-normal stars at similar \feh\ implies that low-mass stars could form at that metallicity, but that masses $M \lesssim 1.4$ \msun\ were disfavored in the IMF. This result is also explained if the abundance-derived top-heavy IMF for primordial stars persists to [Fe/H] $\sim -5.5$. This finding indicates that low-mass star formation was possible at extremely low metallicity, and that the typical stellar mass may have had a complex dependence on metallicity rather than a sharp transition driven solely by gas cooling.
\end{abstract} 

\section{Introduction}

Metals are a major influence on star formation and evolution. By controlling the rate at which star-forming gas can radiatively cool, metals help determine the typical stellar mass.  At one extreme of metallicity, the first stars likely had many unique properties because of their primordial composition. Theoretical studies of the formation of ``Population III'' stars have converged on the clear prediction that their typical Jeans mass scale was $\sim 100$ \msun\ (Bromm \& Larson 2004) because primordial gas with $T \lesssim 10^4$ K was restricted to inefficient cooling by molecular hydrogen formed by slow gas-phase reactions in the absence of dust. This expectation has now been verified with high-fidelity numerical simulations that track pre-stellar evolution until formation of a dense molecular protostellar core (Abel, Bryan, \& Norman 2002; Bromm, Coppi, \& Larson 1999, 2002; Omukai et al. 2005; O'Shea et al. 2006, Yoshida et al. 2006). While radiative feedback during collapse to the main sequence may limit the final mass (Tan \& McKee 2004), massive stars are still expected.

The transition from massive primordial stars to the first low-mass stars is of great interest for understanding the formation of the first galaxies and of the Milky Way. The concept of a unique cooling regime leading to massive stars at low metallicity contrasted with the prevalence of low-mass stars (LMS) in the modern Universe implies that metallicity plays an important role in establishing the conditions for low-mass star formation. Though there need not be a single-valued metallicity criterion for LMS formation, such a sharp ``critical metallicity'', $Z_{crit}$, can be calculated for typical interstellar conditions and used to inform more detailed models.  Bromm \& Loeb (2003) estimated $Z_{crit} \sim 10^{-3.5}$ $Z_{\odot}$ accounting for only the dominant cooling lines of C and O. Schneider et al. (2002) included the cooling effects of dust formed in SNe from primordial stars and found $Z_{crit} \sim 10^{-5.5} Z_{\odot}$. Santoro \& Shull (2006) studied detailed thermodynamic trajectories and more realistic mixtures of relative abundances, and found the critical metallicity to be a function of gas density with characteristic values $Z_{crit} = 10^{-3.5} Z_{\odot}$. Finally, detailed work on metallicity-dependent cooling and star formation by Omukai et al. (2005) and Smith \& Sigurdsson (2007) has suggested that there may be a more gradual evolution, with a possibly bimodal distribution near the transitional metallicity owing to the varying influence of the global thermal environment (Larson 2005). Thus the theoretical picture of the variation of the IMF with metallicity is somewhat unclear and more observational information is needed.

Recent efforts to understand the formation of stars and galaxies in the early Universe using local, long-lived signatures, an approach termed ``near-field cosmology'' (Freeman \& Bland-Hawthorn 2002), have begun to bring insight into the nature of the early stellar generations. Coordinated observing campaigns to discover and study the most metal-poor stars have revealed large samples of stars with \feh\ $< -2$ and abundance patterns that change in character from higher metallicity (Beers \& Christlieb 2005). Attempts to turn these high-quality abundance data on metal-poor stars into empirical limits on the primordial IMF have favored stars with $\sim 10 - 50$ \msun\ as the progenitors of the metals in these presumably second-generation stars (Umeda \& Nomoto 2003, 2005; Tumlinson, Venkatesan, \& Shull 2004; Tumlinson 2006, hereafter T06). This chemical evidence is presently the sole observational test of theoretical models for primordial star formation. The data is in broad agreement with the theoretical expectations for massive stars; more detailed theoretical calculations are needed to better understand the role of feedback (and metals) in limiting the final stellar mass. Because these chemical signatures are borne by LMS, the conditions under which they form is of fundamental importance to their discovery and interpretation. It is therefore important to pursue empirical tests of the IMF at the lowest metallicities to study the implications of low metal abundance on the star formation process. Unfortunately the known populations of metal-poor stars are not yet large enough to determine the critical metallicity empirically (T06; Salvadori, Schneider, \& Ferrara 2007), and the metallicity distribution function (MDF) shows no feature that indicates a sharp increase in the relative incidence of low-mass stars.

The two most iron-poor stars, HE0107-5240 with \feh\ = $-5.3$ (Christlieb et al. 2002, 2004; Bessell et al. 2004) and HE1327-2326 with \feh\ $= -5.4$ (Frebel et al. 2005; Aoki et al. 2006, hereafter AF06) are a surprising development in the ongoing study of ancient stellar ``fossils'', because of their conspicuous presence beyond an apparent gap in the halo metallicity distribution function (MDF) and their extreme relative enrichments of light elements, [C/Fe] $\sim 4$ and [C/H] $\sim -1.5$. Their implications for LMS formation at low metallicity hinge on the interpretation of these enhancements: if their surface abundances exist throughout the entire mass of the star, they would support the view that it is carbon that sets $Z_{crit}$ and that LMS formation is inhibited below [C/H] $\simeq -3$ (Bromm \& Loeb 2003; Frebel, Johnson, \& Bromm 2007). However, the HMPs may have formed with total metallicity $Z \sim 10^{-5} Z_{\odot}$ and acquired surface C enhancement later by, e.g. accretion and/or internal enrichment. If so, these stars imply that LMS star formation was possible, if rare, at that extremely low metal abundance. This paper uses the comprehensive observational information now known about the two hyper-metal-poor stars (HMPs; \feh\ $< -5$, in the terminology of Beers \& Christlieb 2005) to test models for their origin and to draw implications from each model for the IMF at low metallicity.

\begin{figure*}[!ht]
\centerline{\epsfxsize=\hsize{\epsfbox{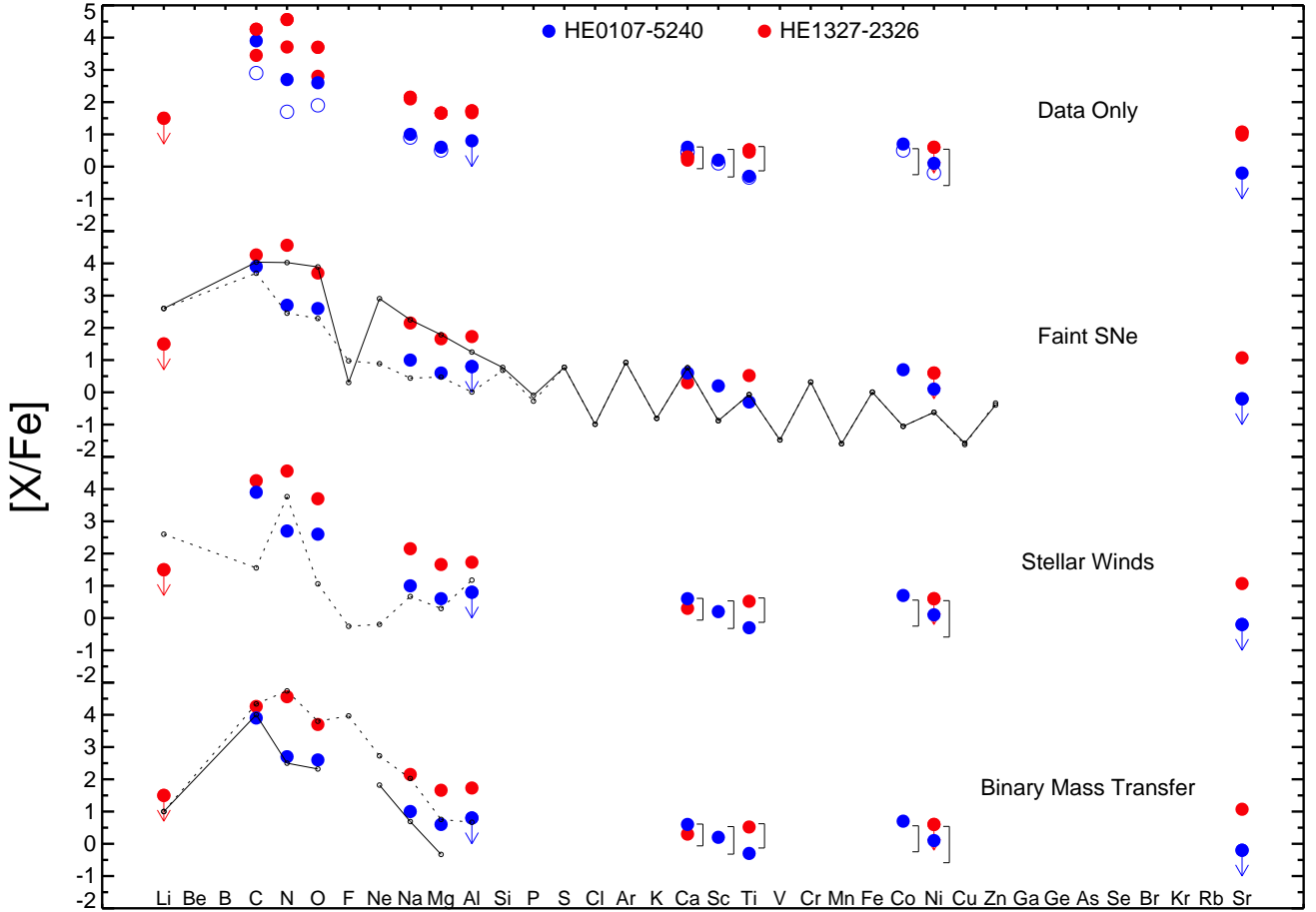}}}
\caption{A comparison of HE0107-5240 and HE1327-2326 abundance patterns with proposed models. The top panel shows the data only, compiled from observational papers by Christlieb et al. (2004), Frebel et al. (2005), and AF06, assuming 1D stellar atmospheres. Open circles mark recent 3d-corrected abundances from Collet et al. (2006). The value plotted for Li is $\epsilon_{Li} \leq 1.5$, not [Li/Fe] $\leq 6$. The lower three panels show proposed models for the abundance patterns, as described in \S~3. Second from top: The faint supernova models constructed by Iwamoto et al. (2005) to match the patterns of HE0107-5240 (dotted) and HE1327-2326 (solid). Third from top: The massive stellar winds model of Meynet, Ekstr\"{o}m, \& Maeder (2006). Bottom: The AGB/binary models of Suda et al. (2004, solid) and Meynet, Ekstrom, \& Maeder (2005, dotted); these models are intended to apply to either star. In the lower two panels, the range bars for Ca, Sc, Ti, Co, and Ni mark the symmetric $\pm 3\sigma$ range occupied by mainstream EMP stars with \feh\ $= -3.5$ to $-2.0$ from Barklem et al. (2005), to represent the plausible range of these elemental abundances if the HMPs have similar stellar progenitors in these elements. \label{hmp-abund-fig}} \end{figure*}

Section 2 of this paper describes the key observed features of the known HMPs. Section 3 reviews and critiques the three main hypotheses that have been proposed to explain the origin of the HMPs, and argues that a binary origin followed by mass transfer from an AGB companion best fits the available data. Section 4 pursues the implications of this finding for the mass function of stars at low metallicity. Section 5 summarizes the conclusions and comments on related issues.

\section{Features of The Hyper-metal-poor Stars}

Since the discovery of HE0107-5240 and HE1327-2326 substantial effort has gone into studying their spectra at high resolution, into modeling their atmospheres to deduce abundances, and into theoretical explanations of their origin. This section briefly reviews the major observed features of these stars before evaluating the theoretical explanations against these data (\S~3). AF06 also provide a thorough review of the abundance patterns of HE0107-5240 and HE1327-2326 from a more detailed observational point of view.

A summary view of the abundances in the two HMPs is shown in Figure~\ref{hmp-abund-fig}, together with model abundance patterns for the hypothetical origins discussed in \S~3. The two abundance patterns are similar in their gross features, but also differ significantly in important elements. The most notable features of the HMP stars are the strong enhancements of the light elements C, N, and O in both stars, and the enhancement of Na, Mg, and Al in HE1327-2326. AF06 report [C,N,O/Fe] = $+4.26, +4.56, <4.0$ for HE1327-2326, along with [Na,Mg,Al/Fe] = $2.55, 1.76, 1.33$. Christlieb et al. (2003) report [C,N,O/Fe] = $+3.9, +2.7, +2.6$ for HE0107-5240. These abundances are adopted here for tests of theoretical models and appear in Figure ~\ref{hmp-abund-fig}. These abundances carry typical errors of $0.05 - 0.2$ dex with the larger uncertainties for the CNO abundances derived from molecular absorption bands.

There is no reason to believe that the observational selections that led to the discovery of the two HMPs introduced a bias in favor of C-enhanced stars, or against C-normal stars. Both stars were selected for spectroscopic followup from the Hamburg-ESO objective prism survey (Wisotzki et al. 2000) based on the strength of their Ca II K bands, which are used as a proxy for \feh. Both stars lie in the mainstream of [Ca/Fe] for metal-poor stars in the halo (Cayrel et al. 2004; Barklem et al. 2005), so there is no reason to suspect that this selection criterion preferentially selects C-rich stars. Finally, the C-bearing molecular bands were not known until high-resolution spectra were obtained (AF06), so there is no reason to suspect that bias was introduced during selection by the investigators. We can therefore proceed under the assumption that the two HMPs are a fair representation of low-mass stars at their epoch and/or metallicity, until proven otherwise.

Another notable feature of HE1327-2326 is its strong apparent depletion of lithium, $\epsilon$(Li) $< 1.5$ (AF06)\footnote{This is the tightest published limit. On the basis of further observations A. Frebel (2006, priv. comm.) now reports $\epsilon$(Li) $< 1.0$; see http://www.int.washington.edu/talks/WorkShops/.}. This limit is well below the primordial abundance inferred for primordial gas from measurements of the cosmic microwave background (CMB) and the theory of Big Bang nucleosynthesis, $\epsilon$(Li) $= 2.6$ (Spergel et al. 2006). HE1327-2326 also lies well below the modestly depleted ``Spite plateau'' measured in other metal-poor stars, for which $\epsilon$(Li) $\sim 2.0 - 2.2$ is typical (Ryan et al. 2002). This upper limit indicates strong lithium depletion in the atmosphere of HE1327-2326, which is surprising for a dwarf or subgiant star of $T_{eff} \simeq 6000$ K.  Although the interpretation of Li is complicated for the giant star HE0107-5240 that may have depleted its own Li during evolution into a giant, the status of Li as a primordial element that is easily destroyed in stars can be used to constrain the possible origin scenarios for the HMPs.

Finally, there is a clear detection of the neutron-capture element strontium ([Sr/Fe] = 1.1) in HE1327-2326, while barium is not detected ([Ba/Fe] $< 1.46$). Sr has multiple nucleosynthetic sources that produce Sr in distinctive ratios to Ba, so it is possible in principle to use the limit on the ratio of these elements, [Sr/Ba] $ > -0.4$, to constrain origin scenarios for HE1327-2326.

The main s-process associated with AGB stars produces a ratio [Sr/Ba] $\lesssim -1$ (Aoki et al. 2002; Barklem et al. 2005) and has a strong metallicity dependence because it requires heavy seed nuclei (Travaglio et al. 2004), even though its neutron source reaction ($^{13}$C($\alpha$, n)$^{16}$O) is primary. Similarly, the weak s-process associated with He-burning massive stars has a strong metallicity dependence through the neutron source reaction $^{22}$Ne($\alpha$, n)$^{25}$Mg. The novel proton-mixing-driven mechanism proposed by Goriely \& Siess (2001) yields a broad range of s-process elements even for zero-metallicity stars, but its intrinsic ratio [Sr/Ba] $\simeq -1.2$ is also incompatible with observed limit in HE1327-2326. Thus the Sr and Ba abundances, whatever their site of production, cannot be dominated by yields from the known slow neutron-capture processes.

There are two neutron capture processes associated with massive stars that produce Sr and Ba in a ratio consistent with the limit from HE1327-2326. The main r-process, which is thought to be associated with explosive nucleosynthesis in massive stars (Truran et al. 2002), most likely of $8 - 10$ \msun\ (Wanajo \& Ishimaru 2005), yields [Sr/Ba] $ = -0.15$ according to the solar system residuals method (Arlandini et al. 1999). The ``light element primary process'' (LEPP) isolated to massive stars because of its clear presence at \feh\ $\lesssim -3.0$ (Travaglio et al. 2004) can give [Sr/Ba] $= 0.0 - 1.0$. The limit on [Sr/Ba] $> -0.4$ in HE1327-2326 is also consistent with the ratio [Sr/Ba] $= -0.5$ to $+0.8$ observed in stars of [Fe/H] $\lesssim -1.5$ for which the main r-process apparently dominates the abundances of Eu and Ba (Barklem et al. 2005). The observed scatter in [Sr/Ba] in this sample, across the full range of the main r and LEPP values, probably arises in decoupling between the sources of the main r-process and the LEPP. This scatter makes them difficult to distinguish in individual cases. The Sr production that preceded HE1327-2326 therefore could be dominated by either, or a combination, of these processes. It may even contain a small contribution from the s-process sites above, but this contribution cannot dominate the observed abundances. Unfortunately, the ambiguous origin of Sr in stars of [Fe/H] $\lesssim -2$ prevents decisive tests of HMP origin scenarios using this single ratio (see \S~4).

The non-detection of Sr in HE0107-5240, with [Sr/Fe] $< -0.2$, is consistent with the high degree of scatter seen in this ratio for metal-poor stars in the HERES survey (Barklem et al. 2005), and does not prefer a specific nucleosynthetic origin.

\section{Proposed Origins for HE0107-5240 and HE1327-2326}

The goal of this paper is to select from among the existing hypotheses a common origin scenario for HE0107-5240 and HE1327-2326 that can also account for their differences. This section examines proposed hypotheses for the origin of the two HMPs, including the faint supernova  hypothesis of Umeda \& Nomoto (2003) and Iwamoto et al. (2005), the massive star winds model of Meynet, Ekstr\"{o}m, \& Maeder (2006), and the Pop III binary hypothesis of Suda et al. (2004).

\subsection{Self-enrichment of CNO?}

Production of CNO {\em in situ} does not appear possible for the two HMPs. First, HE1327-2326 is an unevolved star on either the main sequence or subgiant branch of the HR diagram, and there is no known mechanism for strongly enhanced primary light element enrichment in such a star. AF06 state that {\em in situ} processes cannot account for the N excesses of HE1327-2326 and HE0107-5240.

\subsection{The ``Faint Supernova'' Hypothesis}

Using a sophisticated stellar evolution and nucleosynthesis code, the Tokyo group has calculated explosion models and nucleosynthetic yields for metal-free stars with a wide range of explosion energies and up-to-date physics. Their general approach has been to examine the wide range of observed supernova types at low redshift, with varying energy inferred from optical spectra, and ask what abundances would be expected in metal-poor stars if the same wide variation in explosion energy, $E \equiv 10^{51} E_{51}$ erg, from $E_{51} < 1$ to $E_{51} = 100$, occurred in primordial stars. This approach has been fruitful, as it has provided possible explanations for the unusual trends in iron group abundances (e.g., Cr, Mn, Co, Zn) in EMP stars (Tominaga et al. 2006) and also provides a novel model for the origin of HE0107-5240 and HE1327-2326. This critique of the hypothesis focuses on the later paper by Iwamoto et al. (2005), who model the abundance patterns for both HMP stars.

In papers by Umeda \& Nomoto (2003) and Iwamoto et al. (2005), the Tokyo group has presented models for ``faint supernovae'' that explode with $E_{51} < 1$ and undergo mixing of material across the ``mass cut'' dividing ejecta from material that falls back into the black hole. Their models for HE0107-5240 and HE1327-2326 both have 25 \msun\ and differ only in their explosion energies and mixing-fallback parameters. For HE0107-5240, they propose a model with $E_{51} = 0.71$ and for HE1327-2326 they propose $E_{51} = 0.74$. From the point of view of the full range of SN energies explored in other models from the same group ($E_{51} = 1 - 100$), these two explosion energies are quite similar but nevertheless reproduce the very different abundance patterns of the two HMPs very well, including the significantly higher ratios of N, Na, Mg, and Al relative to Fe seen in HE1327-2326 (Figure~\ref{hmp-abund-fig}). Their models introduce homogeneous mixing in the entire region between the compact core and the mass cut (determined to be 6.3 and 5.8 \msun\ for HE0107-5240 and HE1327-2326, respectively, from a derived $E_{51}-M_{cut}$ relation), followed by the ejection of all material outside the mass cut and a fraction, $f$, of the material from inside the mass cut. This latter assumption, which introduces the additional free parameter $f$, is necessary for the escape of heavier iron-group elements from the dense region near the core that would otherwise fall back to the remnant. Thus the very different and unusual abundance patterns of the two HMPs can be explained self-consistently by a single physical mechanism associated with massive stars.

The faint SN hypothesis makes three specific predictions: first, that the expected continuous variation in $E_{51}$ and therefore in ejected Fe mass causes continuous variation in \feh. The apparent gap between the HMPs and the mainstream EMP stars in the halo is not consistent with this prediction, so Iwamoto et al. (2005) suggest that ``jet-induced mixing might be responsible for constraining the distribution of the $f$ value''. Second, assuming that the formation of low-mass stars requires a minimum critical metallicity, $Z_{crit}$, the faint SN model implies increasing [C/Fe] with decreasing [Fe/H]. This effect is seen in the available data on a few stars (see Figure~13 of AF06 and Frebel et al. 2007), but since this prediction is generic to any single-progenitor explanation for the HMPs it would not uniquely favor the faint SN model. Third, since variations in [Na, Mg, Al/Fe] are caused by variation in $E_{51}$, these ratios should show a continuous distribution. Unfortunately this prediction cannot be tested by further observations of these two stars and must await the discovery of more HMP stars.

However, substantial fine tuning of the faint SN model is required to match the observed abundances. Though the progenitor mass is held fixed, the explosion energy for the two favored models differs by only 4\%. The diffusion coefficient for mixing between the He convective shell and H-rich envelope must be increased by a factor of $30$ increase to reproduce the high N/C ratio seen in HE1327-2326, with no apparent external justification.

The faint SN model also does not explain the presence of Sr and the limit on Li in HE1327-2326. The Sr detection with [Sr/Ba] $> -0.4$ could be produced by a massive star with either the main r-process (Truran et al. 2002; Wanajo \& Ishimaru 2005) or the newly discovered light neutron-capture element primary process thought to be responsible for Sr-Y-Zr in the mainstream EMP halo stars (Travaglio et al. 2004). Though the 25 \msun\ faint SNe are possibly capable of Sr nucleosynthesis, the models have not yet been extended to show that the proper conditions are realized, though the Sr/Ba ratio favors massive stars over s-process origin in an AGB star.

The main difficulty encountered by the faint SN model is the strong depletion of Li in HE1327-2326, $\epsilon$(Li) $<1.5$. This abundance lies a full decade below the WMAP cosmological value of $\epsilon$(Li) $= 2.6$ (Spergel et al. 2006) and substantially below the ``Spite plateau'' with $\epsilon$(Li) $ = 2.0 - 2.2$ over \feh\ $= -2.5$ to $-4.0$ (Spite \& Spite 1982; Ryan et al. 2001). The 25 \msun\ fiducial faint SN models eject exactly 0.20 and 0.12 \msun\ of C for HE1327-2326 and HE0107-5240, respectively, so the mass of fresh gas into which the ejecta must be mixed to obtain the observed C abundance relative to H is $\lesssim 1000$ \msun\ -- smaller than the $\sim 10^4$ \msun\ expected for SNe with $E_{51} \sim 1$. The strict requirement that the ejecta of the fiducial faint supernovae mix with exactly 650 \msun\ to match the observed [C/H] $=-1.14$ in HE1327-2326 (Iwamoto et al. 2005) entails a recovery to within 4\% of the primordial or plateau Li abundance even after complete astration in the 25 \msun\ progenitor. If the massive star has astrated less than the total abundance of lithium with which it formed, the diluted ejecta will lie even closer to the cosmic value, so this assumption of complete astration is a conservative one. Thus, the faint SN model cannot explain the observed ratios [C/H], [C/Fe], and the limit on Li simultaneously. This line of reasoning for Li applies as well to the general class of models in which a ejecta from a single primordial supernova is diluted into a large reservoir of primordial gas prior to the formation of HE1327-2326.

The faint SN hypothesis provides a reasonable, physically motivated model for the abundances in the HMPs, but it cannot yet explain the full range of observed properties of these stars. Clearly this hypothesis cannot yet be firmly excluded, but it cannot be considered complete until it can explain the Li depletion in HE1327-2326 and until the degree of fine-tuning needed to make it work has been reduced. If it is eventually proven correct, and when combined with the apparent absence of C-normal stars at \feh\ $=-5$, it would corroborate the theoretical idea that the critical metallicity is determined mainly by cooling radiation of C and O, such that $Z_{crit} \sim 10^{-3} Z_{\odot}$ (Bromm \& Loeb 2003; Santoro \& Shull 2006; Frebel et al. 2007).

\subsection{The Massive Stellar Winds Hypothesis}

Meynet, Ekstr\"{o}m, \& Maeder 2006, hereafter MEM06) and Karlsson (2006) have proposed that the HMP stars formed with \feh\ $\simeq -5.5$ and [C/Fe] $\simeq +4$ and acquired their light-element enhancements prior to formation from CNO-rich winds driven by rapidly rotating massive stars. MEM06 construct rotating stellar models with the key assumption that the total angular momentum with which a star forms is determined by its mass, invariant with metallicity. This simple assumption leads to $v = 800$ \kms\ at the surface of a 60 \msun\ star, which rotates near breakup velocity throughout its main-sequence lifetime. Deep mixing currents driven by this fast rotation bring light elements produced in the core up to the surface where fast rotation sends them off in a wind. The ejecta are extremely enhanced in CNO, and according to MEM06 resemble the HMPs in their overall composition (see Figure~\ref{hmp-abund-fig}).

While the case for fast stellar rotation at very low metallicities is favored on other grounds (see Chiappini et al. 2006 on the ratios of N/C and N/O seen in unevolved stars of \feh\ $\lesssim -3$), it appears that the specific rotating massive star models proposed by MEM06 do not reproduce all the observed features of HE0107-5240 and HE1327-2326. As shown in Figure~\ref{hmp-abund-fig}, the MEM06 winds-only model predicts ratios of N/C and N/O that are $1 - 2$ dex too high to match the data for the HMPs. Models that include the light-element yields of the entire star (intended to represent the yields of a supernova) and which match [O/Fe] by construction can improve the fit to [C/Fe] but not [N/Fe] (see Figure 9 of MEM06).

The rapidly rotating massive stars model, like the faint supernova model, has trouble meeting the strong upper limit on the abundance of Li in HE1327-2326, $\epsilon$(Li) $ < 1.5$. The reason is that the calculated yields of light elements from the winds and SN require dilution into $\sim 10^5$ \msun\ of primordial interstellar gas to give the observed [Fe/H] $= -5.4$ and [O/Fe] $ = 3.5$ for HE1327-2326. Indeed, in constructing the observed abundance ratios MEM06 calculate that the mass of the total ejecta from the star, $m_{ej} \simeq 50$ \msun, represents less than 1 part in 4000 of the total mixing (or dilution mass), from which HE1327-2326 eventually formed. If, as in the MEM06 scenario, the HMPs had only one massive star progenitor, the Li abundance in the diluted material from which the HMPs form should have been reduced by the same ratio, 1 part in 4000, instead of the factor of more than 5$-$10 that is apparent in the data.

Piau et al. (2006) have found that the lithium depletion seen in EMP stars could be caused by astration in early generations of massive stars, but if so the total mass formed into stars, and therefore astrated, must be a large fraction of the material into which their ejecta are eventually mixed before formation of the low-mass second generation. This avenue appears to be closed for the MEM06 model, since the progenitors of the HMPs must be an individual star of zero or extremely low metallicity and with a large dilution mass. It is also difficult to see how the low Li abundance and low dilution mass could be reconciled with the low Fe abundance if many primordial stars have contributed. Thus, the generic class of models that requires high dilution of massive star ejecta into primordial material cannot account for the strong lithium depletion observed in HE1327-2326. This line of reasoning applies equally well to the model proposed by Chieffi \& Limongi (2003), in which HE0107-5240 arose from two primordial SNe that mixed their ejecta together before forming HE0107-5240.

\subsection{Binary Mass-Transfer Hypothesis}

The binary mass-transfer hypothesis proposes that the HMP stars acquired their light element enhancements only at their surfaces from a companion that passed through an AGB phase. This hypothesis was raised by AF06 and MEM06, but it has been most fully developed by Suda et al.~(2004) for HE0107-5240 and Komiya et al. (2007) for the general class of carbon-enhanced metal-poor stars at [Fe/H] $\lesssim -2$. They present new models of light-element and neutron-capture nucleosynthesis in low-metallicity stars of $0.8 - 8$ \msun\ that evolve through an AGB phase and experience deep mixing in their interiors, accompanied by increased production and dredge-up of C, N, O, Na, and Mg.  Suda et al. (2004) find that primary stars of $M_1 = 1.2 - 3$ \msun\ experience a series of $20 - 30$ deep mixing events driven by He-flash burning in the interior (helium-flash driven deep mixing; He-FDDM). These dredge-up events bring to the surface the products of core nucleosynthesis, which can then escape in a wind and be captured by the low-mass partner ($M_2 \sim 0.8$ \msun). In this model, the dredge-up events are directly responsible for the C and N enrichments in the HMPs. Once $^{13}$C has been mixed up into the convective zone, a series of $\alpha$- and neutron-capture reactions produce O, Na, and Mg. In low-metallicity stars of $1.2 - 3$ \msun, the chemical composition of the outer layers at the end of He-flash burning quantitatively match the abundances of HE0107-5240, including a predicted carbon isotope ratio $^{12}$C/$^{13}$C $= 30 - 120$ that corresponds well with the observed ratio of $ > 50$ (Christlieb et al. 2004). Suda et al. (2004) argue that binary partners of $M < 1.2$ \msun\ fail to explain the observed C/N ratio of HE0107-5240, while stars above 3 \msun\ do not experience the He-FDDM phase (Komiya et al. 2007 revise this limit to 3.5 \msun, which is adopted below in \S~4). The specific abundance ratios also constrain the initial separation of the binary pair and mass loss considerations allow Suda et al. (2004) to calculate that the HE0107-5240 pair should presently show a period of approximately 150 yr and a radial velocity of $7$ km s$^{-1}$, an orbit too wide and slow to have been detected by radial velocity monitoring to date.

The most important aspect of the binary mass-transfer hypothesis is that it can naturally explain the observed limit on the abundance of lithium in HE1327-2326. In this scenario, the mass transferred from the primary to the secondary has been dredged from hotter layers in the interior of the primary and is therefore devoid of lithium, which burns at $\sim 2.5 \times 10^6$ K. When this material mixes with the undepleted outer layers of the unevolved secondary, the surface lithium abundance drops below the primordial value. Based on the CNO ratios, Suda et al. (2004) estimate that a total mass of order $M_{acc} \simeq 0.01$ \msun\ was transferred from the primary to HE0107-5240. This mass is of the same order as that estimated by Norris et al. (1997) for the mass that could lead to $\epsilon$(Li) $\simeq 1.0$ in the surface convection layer of a secondary with $M_2 \sim 0.8$.

The Li depletion, C/N ratio, and C isotope ratio in HE1327-2326 can be used to refine the details of the binary-mass transfer model for HE1327-2326. The Li depletion implies that the surface convective zone was diluted by a factor of $4 - 15$ by the accreted material, so $M_{acc} = 0.012 - 0.15$ \msun\ for $M_{conv} = 0.003 - 0.01$ \msun\ (Figure 1 here depicts $\epsilon$(Li) $= 1.0$, corresponding roughly to $\Delta M = 0.01$ \msun). Because the accreted material likely dominates the ``dilution mass'' in the convective zone, the C abundance in the accreted material was close to the observed value. The $4-15\times$ dilution then implies that the HE1327-2326 primary had [C/H] $\sim -2.2$ to $-1.5$ at its surface prior to the mass transfer. According to Suda et al. (2004), HMP stars exit the He-FDDM phase with surface abundances [C/H] $\sim -2.5$ and C/N $\sim 5$. Increases above those values are caused by the mixing of $^{12}$C to the surface by successive ``third dredge-up'' (TDU) episodes. In comparison with HE0107-5240, HE1327-2326 has a lower C/N ratio, which together with its low Li abundance implies that it experienced few, if any, TDU episodes that raised the surface C abundance above [C/H] $\sim -2.5$. This suggestion can be cross-checked with C isotope ratios, since TDU brings only $^{12}$C to the surface. Thus HE1327-2326 would be expected to show a lower $^{12}$C/$^{13}$C than HE0107-5240, where Christlieb et al. (2004) find $^{12}$C/$^{13}$C $> 50$. AF06 report $^{12}$C/$^{13}$C $> 5$ for HE1327-2326, but the data quality is too low to obtain a measurement. This is consistent with the finding of fewer TDU episodes, and a better dataset will eventually allow a test of this prediction. In summary, the Li abundance of HE1327-2326 can easily be accommodated if a few hundredths of a solar mass with [C/H] $\sim -1.5$ moved from a primary to the secondary after a He-FDDM event and a few TDU episodes. This model also explains the high N/C ratio of HE1327-2326 more naturally than the faint SN model, which must arbitrarily invoke extra diffusive mixing between the H and He burning layers that is not applied to HE0107-5240.

The presence of Sr in HE1327-2326 may indicate the mass of the primary star, but the signal is ambiguous. As detailed in \S~2, [Sr/Ba] $>-0.4$ resembles the r-process or LEPP in massive stars, which give [Sr/Ba] $\simeq -0.5 - 0.8$. The observed ratio rules out a dominant contribution from the main s-process and its close relatives, for which [Sr/Ba] $\sim -1$ (see \S~2). If interpreted literally, this restriction implies that the primary had $M_1 = 0.8 - 1.2$ \msun, where Komiya et al. (2007; case I) find that stars of [Fe/H] $\lesssim -4.5$ do not dredge s-process elements up to their surfaces. If the HE1327-2326 primary had 1.2 - 3.5 \msun, it would be expected to bring s-process elements to the surface and transfer these to HE1327-2326 along with CNO (case II'). However, even this expectation is uncertain. If the system formed with [Sr/Ba] $\simeq 0.8$, as seen in the r-process dominated EMP stars (Barklem et al. 2005), then $4 - 15\times$ dilution of the surface convective zone by s-process rich material would bring [Sr/Ba] down to $\simeq -0.4 - 0.1$, still consistent with the observed limit. While most C-enhanced EMP stars with s-process enhancement (the CEMP-s stars; Barklem et al. 2005; Jonsell et al. 2006) have a ratio [Sr/Ba] $\simeq -1$ that resembles the main s-process, there are a few that scatter up to [Sr/Ba] $\simeq -0.4$, marginally consistent with the limit in HE1327-2326. These results indicate that the metallicity dependence of Sr nucleosynthesis is not well understood, even in stars where the C-enhancement is clearly the result of a binary mass transfer event. If the process that acted in these other C- and s-process enriched stars also acted in the primary of HE1327-2326, it could have had 1.2 - 3.5 \msun (case II', Komiya et al. 2007) and still reproduce the observed abundances. Thus, the primary mass of HE1327-2326 cannot yet be conclusively constrained any further than the range 0.8 - 3.5 \msun, though improved limits on neutron-capture element abundances can resolve this ambiguity.

The binary mass-transfer hypothesis has strong empirical justification in other Pop II samples. Ryan et al. (2002) established an observational link between lithium depletion, binarity, and mass transfer in metal-poor stars. They examined four stars for which lithium had not been detected, with limits similar to that for HE1327-2326. Of the four, three are confirmed binaries and show fast rotation consistent with mass transfer between the partners of order $M_{acc} = 0.001 - 0.01$ \msun. Thus the link between binarity and mass transfer has been established for stars that share the characteristic of strong lithium depletion with HE1327-2326. The major lacking element in the HMPs is a clear detection of radial velocity variations at the $\sim 10$ km s$^{-1}$ level predicted by Suda et al. (2004).

The present model modifies the Suda et al. (2004) picture, which assumed that HE0107-5240 and its binary companion were primordial (Pop III) stars that acquired all their heavy elements by accretion after formation. This assumption is not necessary to describe their abundances. In the present model, the Fe-peak elements, some of the $\alpha$ elements, and Sr were formed into the HMPs and the light elements were acquired from the binary companion, such that the two HMPs were originally binary Pop II stars. Thus the relative abundances of these elements, which should be undisturbed by the mass transfer event, should resemble other metal-poor stars that descended from the same early generations of stars. To test this idea, Figure~\ref{hmp-abund-fig} compares Ca, Sc, Ti, Co, and Ni abundances for the HERES sample of Barklem et al. (2005) to the HMPs. The plotted range bars next to the HMP data points for these elements show the symmetric $\pm 3 \sigma$ range from the mean for these 253 stars with \feh\ = $-3.5$ to $-1.5$. The relative abundances in the mainstream EMP stars also describe the HMPs; the match is somewhat better than the faint SN model for Sc, Co, and Ni. This comparison corroborates the idea that the HMP stars could have acquired their Fe-peak abundances from the same general population of primordial and/or low-metallicity stars that provide the early enrichment of these elements in the mainstream EMPs, and need not have originated in some exotic mechanism.

MEM06 also consider the binary mass-transfer hypothesis, on the grounds that the evolution and nucleosynthesis in intermediate-mass stars near 8 \msun\ are not substantially different from the massive stars on which their abundance models are based. They present a 7 \msun\ model with rapid rotation (800 km s$^{-1}$), and show that the abundances in the stellar envelope at the end of core He-burning more closely match the N/C and N/O ratios of the HMPs than their 60 \msun\ model, and can also give $^{12}$C/$^{13}$C $=20 - 2500$ (see their Figure 10, and the dotted line in the bottom panel of Figure~\ref{hmp-abund-fig} here). MEM06 state that the [Sr/Ba] $> -0.4$ limit in HE1327-2326 cannot be reproduced by the main s-process in the intermediate mass primary, and that the massive star winds hypothesis is preferred on these grounds. However, it is not required that the Sr and the CNO come from the same source, as the observed [Sr/Ba] ratio is consistent with neutron-capture nucleosynthesis in preceding massive stars and so could have formed with the HMPs. It need not have come in from the mass-transfer and so does not affect the case for the binary mass-transfer hypothesis.

A measurement or improved limit on Ba in HE1327-2326 could help resolve the ambiguity in neutron-capture origins and assess the viability of the model proposed by Suda et al. (2004) in which the HMPs acquire their Fe-peak elements by accretion from the interstellar medium. In this model, the star should spend most of its time in the higher metallicity regime during this time, \feh\ $\gtrsim -2$, and therefore to acquire neutron capture abundances that reflect the mainstream ratios seen in metal-poor stars at this metallicity. If so, the observed ratio should move away from the observed low-metallicity ratio, [Sr/Ba] $\gtrsim -0.5$, and toward the pure s-process value [Sr/Ba] $\lesssim -1$ (Barklem et al. 2005). This model is already marginally inconsistent with the data, and makes it more likely that HE1327-2326 acquired its Sr at formation from a prior supernova, which could scatter [Sr/Ba] up as far as $\sim 1$. Further tightening of the [Sr/Ba] limit, or a Ba detection, could bolster this constraint and possibly further constrain the mass of the primary star.

The main failing of the mass-transfer binary model is that it does not perfectly match the abundances of Mg and Al, particularly in HE1327-2326. Suda et al. (2004) suggest that there might be neutron-capture pathways for Mg production, but it is not yet clear that this mechanism can reproduce in the proper ratios. Indeed, the faint SN and massive star winds reproduce [Al/Fe] better, but all three fall short of the strong Al enhancement in HE1327-2326. Fenner et al. (2004) present yields of Na, Mg, and Al for AGB stars that better match the observed abundances relative to Fe, and which could improve the fit if they can be integrated with the Suda et al. (2004) model for C-enrichment. This slight disagreement does not disfavor the binary mass-transfer hypothesis in light of the other patterns seen above.

In summary, the binary mass-transfer hypothesis proposes that HE0107-5240 acquired its light-element enrichment from a primary of mass 1.2 - 3 \msun\ (Suda et al. 2004), while the C, N, Li, and Sr abundances of HE1327-2326 indicate a primary of 0.8 - 3.5 \msun, with lower masses required if HE1327-2326 accreted no s-process Sr. Binary mass-transfer is the only scenario evaluated here that can meet the strict upper limit on Li in HE1327-2326. For these reasons, it is adopted below for further investigations into the IMF of hyper-metal-poor stars.

\subsection{Comments on the Origin Scenarios}

The foregoing sections evaluated the proposed origin scenarios for HE0107-5240 and HE1327-2326 and argued that only the binary mass-transfer hypothesis adequately explains the available data. For simplicity, this discussion has been limited to considering the origin scenarios one at a time, and not in combination. Clearly, the faint SN and massive star winds models cannot match the available data on their own, but it is possible in principle that the CNO enhancements in the HMPs were acquired from one of the first two mechanisms and the Li depletion was caused by a later binary mass transfer event. In such a picture, C-rich ejecta from massive star winds or faint supernovae enriched the general ISM (Iwamoto et al. 2005; MEM06; Karlsson 2006) from which a binary system formed. Then later a mass transfer across the system transferred Li-depleted materials to the outer layers of the already C-rich secondary to match the constraint on Li seen in HE1327-2326. Combination models of this type are possible and might help explain the diversity of abundance patterns in light, alpha, and neutron capture elements in metal-poor stars. However, each of the three origin scenarios have uncertainties in input physics and parameters that are only magnified by attempting to combine them. It therefore seems wise to restrict consideration to a single model, and combine them only when when new data demands it. For the two HMPs, it is not necessary to make the model more complicated than a binary mass-transfer event in an originally C-normal system.

\section{Implications for the Stellar IMF at \feh\ $ = -5$}

Using detailed comparisons of abundance patterns and proposed models, I have argued that the HMPs formed as the low-mass secondary in a binary system with an intermediate-mass star. In this scenario, the light elements seen in the HMPs come from the AGB primary, and the heavier $\alpha$-capture elements, the iron-group elements, and the neutron capture elements originated in the massive star progenitor(s) that the companions shared. Given the reasonable fit of the mainstream EMP data to the Fe-peak abundances of the HMPs, these progenitors may have been primordial or low-metallicity stars like those that preceded the mainstream EMPs. Having selected this model, we can ask what implications follow for star formation in extremely metal-poor environments.

The first, and perhaps most important implication, is that low-mass stars {\em can} form at $Z \sim 10^{-5.5} Z_{\odot}$, even if only in binary systems. Thus if there is a clear-cut, single metallicity criterion for low-mass star formation (as proposed by, e.g. Bromm \& Loeb 2003), then that metallicity is about this value. More likely, this critical metallicity criterion exists but manifests not as a step function but instead is smoothed out by some other effect, perhaps environmental conditions such as gas density, UV radiation backgrounds, or dust abundance and grain size distributions (Schneider et al. 2002), or detailed relative abundances (Santoro \& Shull 2006).

Another implication of the binary mass-transfer hypothesis follows from considering not only the presence of these two CNO-rich, low-mass companions of intermediate mass stars, but also the apparent absence of isolated, C-normal stars at the same \feh. If the binary star statistics of stars with \feh\ $\sim -5$ are similar to those at solar metallicity, with binary fractions of 60\%, then isolated low-mass stars should easily outnumber pairings with the exact masses (a primary of $1.2 - 3$ \msun\ and a secondary of $\sim 0.8$ \msun\ for the Suda et al. 2004 model) needed to reproduce the observed abundances. If the HMP systems formed with a certain orbital separation and/or eccentricity, then the implied configurations should be still rarer. Yet in an apparently unbiased sample of metal-poor stars, {\em both} of the first two HMPs to be discovered arose in a particular binary configuration that leaves the observed abundance patterns. This leaves the puzzling implication that in the stellar generations with \feh\ $\sim -5$, binary pairings between low- and intermediate-mass stars are more common than isolated low-mass stars.

Theoretical studies of star formation at low and zero metallicity suggest a simple solution to this puzzle that is fully consistent with the observed data: it is that LMS/IMS binaries outnumber isolated LMS because the IMF is top-heavy, or skewed toward IMSs at \feh\ $\sim -5$. For example, an IMF with characteristic mass $\sim 3$ \msun\ and a suitable binary fraction will form relatively few isolated low-mass stars, but if gas fragmentation with metal-line cooling is sufficient then binaries could form with {\em total} mass $\sim 3$ \msun\ and one low-mass secondary companion. After a suitably long time the primary stars will evolve off the main sequence, eject some light elements, and fade to a white dwarf. The secondaries are then seen as low-mass, extremely metal-poor stars with CNO enhanced surfaces, as proposed by Lucatello et al (2005) and Komiya et al. (2007) for the CEMP stars. To reproduce the observed numbers, low-mass stars formed in binaries and fated to experience CNO enrichment must be more common than isolated low mass stars, and a top-heavy IMF is a natural way to do this.

A simple stellar population model can determine what IMF is capable of giving the observed result. A random sample of a hundred stars of solar metallicity in the nearby Galactic neighborhood is rather unlikely to include more than few binaries that consist of a primary with $M_1 = 1.2 - 3.5$ \msun\ and a secondary with $M_2 \lesssim$ 0.8 \msun, the combination that is needed to give rise to systems like HE0107-5240 and HE1327-2326 in the binary hypothesis. The likelihood of this outcome can be calculated precisely for an arbitrary IMF and binary fraction, as follows. The IMF is described by either a broken power law or log-normal function. The log-normal function is:
\begin{equation}
\ln \left( \frac{dN}{d\ln M} \right)= A - \frac{1}{2\sigma ^2}
       \left[ \ln \left( \frac{M}{M_c} \right) \right] ^2
\end{equation}
where $\sigma$ is the width of the distribution, $M_c$ is the characteristic mass, and $A$ is an arbitrary normalization.  This IMF has the advantage of more flexible behavior with only one more parameter than an unbroken power law. The ``normal'' IMF is that given by Kroupa (2002) a power law dependence, $dN/dM \propto M^{-2.3}$ from $0.5 - 140$ \msun, and $dn/dM \propto M^{-0.5}$ from $0.1 - 0.5$ \msun. Both functions run from 0.1 to 140 \msun.

I further assume, in accord with theories for the formation of binary stars in the Galactic neighborhood (Tohline 2002), that most fragmentation into binaries follows the formation of the protostellar core, such that the IMF describes the distribution of ``core masses'', equivalent to the stellar mass for isolated stars or the total mass of binaries. A fraction of the ``cores'', $f_c$, are assumed to be binaries, subject to the condition that a fraction $f_b = 0.6$ (Duquennoy \& Mayor 1991) of all stars with $0.5 - 8$ \msun\ end up in binaries. Triple and higher-order multiple systems represent less than 5\% of the local stellar population (Duquennoy \& Mayor 1991) and are ignored here. Thus $f_c \equiv f_b / (2 - f_b)$. To subdivide ``cores'' with $0.5 - 8$ \msun\ into two stellar masses, $M_1$ and $M_2$, random variates are drawn from the observed mass ratio ($q \equiv M_2 / M_1$) distribution for binaries involving intermediate mass stars, $f(q)$, which is flat from $0.4 - 1.0$ \msun, has a peak near 0.3 \msun, and is fixed to 0.0 at $< 0.1$ \msun\ (Mayor et al. 1992). The binary fraction of extremely metal-poor stars is largely unknown, so conditions like the solar neighborhood must be assumed. Lower binary fractions strengthen the IMF constraints discussed below, and for higher $f_b$ the results are qualitatively similar to $f_b = 0.6$ even for $f_b = 0.9$.

\begin{figure}[!t]
\centerline{\epsfxsize=\hsize{\epsfbox{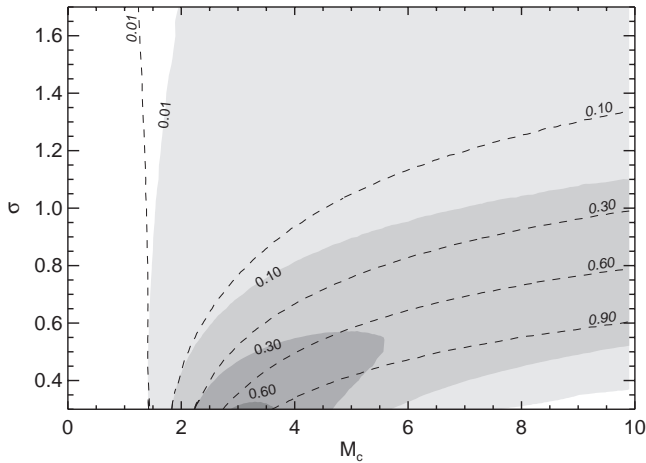}}}
\caption{Constraints on stellar IMF, expressed in terms of the characteristic mass $M_c$ and width $\sigma$ for the log-normal IMF. The grayscale contours mark 1, 10, 30 and 60\% probability of the observed result, from light to dark, for the model in which the HMP primaries had 1.2 - 3.5 \msun. The dashed lines with italic labels show the contours for 1, 10, 30, 60, and 90\% probability where the primaries have 1.2 - 8 \msun. In this latter case, larger $M_c$ are permitted. \label{IMF-prob-fig}} \end{figure}

With these simple assumptions, it is straightforward to calculate the relative probabilities of low-mass stars expected to appear as C-normal HMPs and of binary systems that result in a C-rich HMP star like HE0107-5240 and HE1327-2326. For a normal Galactic IMF (Kroupa 2002), binaries such as give rise to the HMPs are rare, and any unbiased sample that finds two such systems should find many more isolated low-mass stars.  For the Kroupa (2002) IMF and $f_b = 0.6$, a sample containing two binaries with $M_1 = 1.2 - 3.5$ \msun\ and $M_2 = 0.5 - 0.8$ \msun\ should discover $30 \pm 5$ isolated stars with $M = 0.5 - 0.8$. Of course, such a result is not at all expected for a total sample of only a few hundred metal-poor halo stars, since stars with \feh\ $\leq -4$ are intrinsically rare. Instead, we should ask: if two low-mass stars are drawn at random from the Kroupa (2002) IMF, what is the probability of finding that both formed in binaries suitable for producing the observed HMP abundances? This situation is rare, with the probability of obtaining two $0.5 - 0.8$ \msun\ secondary stars of $1.2 - 3.5$ \msun\ primaries, instead of two isolated stars with $0.5 - 0.8$ \msun\ and mainstream [C/Fe], is 0.035\%. If the Galactic IMF is represented by the Miller \& Scalo (1979) function ($M_c = 0.1$, $\sigma = 1.57$), the probability of obtaining the actual result is 0.06\%. These probabilities increase by only a factor of $\sim$1.2 if primary stars of mass up to 8 \msun\ are allowed, such as if all intermediate mass AGB primaries can give the observed HMP abundances. Increasing the binary fraction increases the likelihood of the actual result, but the probability reaches only $\sim$1\% for binary fraction $f_b \gtrsim 0.9$ because most binaries are composed of low-mass pairings, and LMS/IMS binaries are still rare. We can therefore reject at $3\sigma $ the null hypothesis that the solar-neighborhood IMF and binary star statistics describe the stellar populations formed at \feh\ $\simeq -5.5$.

If the solar-neighborhood IMF is excluded by the abundances of the HMPs and the absence of isolated stars at this metallicity, what IMFs are permitted by this unexpected result? To answer this question Figure~\ref{IMF-prob-fig} shows contours of the probability of obtaining the actual result, two appropriate binaries and no C-normal HMPs, given binary fraction $f_b = 0.6$ and assuming either primaries of $M_1 = 1.2 - 3.5$ \msun\ or $M_1 = 1.2 - 8$ \msun\ and a log normal IMF with the parameters as marked on the axes. To obtain even 1\% probability of the observed result, it is necessary to have $M_c \gtrsim 1.4$ \msun. If the HMP primaries must have $1.2 - 3.5$ \msun\ (grey contours), the preferred IMF is narrowly peaked in the range of intermediate mass stars, with $M_c = 2 - 4$ \msun. If more massive primaries are capable of undergoing TDU and so are eligible primaries, then larger $M_c$ are favored, with similar lower limits. These results provide a strong indication that the IMF in gas of the metallicity that formed the HMPs favored stars above a few solar masses and a few times higher than the mean stellar mass $M \sim 0.6$ \msun\ in the Galaxy today.

These results also suggest that in this model, C-enhancement of HMP stars is not guaranteed if the range of primary masses leading to C-enhancement is restricted to only a subset of IMS, or if only a fraction of binaries realize the proper conditions. These systems could appear as C-normal HMPs paired with a white dwarf that never released a C-rich wind, or C-normal isolated HMPs, or as HMPs paired with another C-normal low-mass star. For a given IMF, and with more certain mapping between primary mass ranges and C-enhancement and/or abundance patterns, we can specify the true fraction of HMP stars that should or should not show C-enhancement. Of course, a 100\% C-enhanced fraction is also possible if all LMS at [Fe/H] $\sim -5.5$ are paired with an IMS, and all IMS realize the proper conditions. To resolve these issues, more detailed theoretical work is needed to specify the evolutionary phases and resulting abundance patterns from IMF primaries of different masses. More observational work is needed to improve constraints on abundances, particularly for the neutron-capture elements, for the known stars and to discover new ones that will better fill out the possible range of behavior. These efforts will have sure impact on our understanding of low-mass star formation during the earliest phases of the Milky Way's history.

\section{Discussion and Conclusions}

\begin{figure}[!t]
\centerline{\epsfxsize=\hsize{\epsfbox{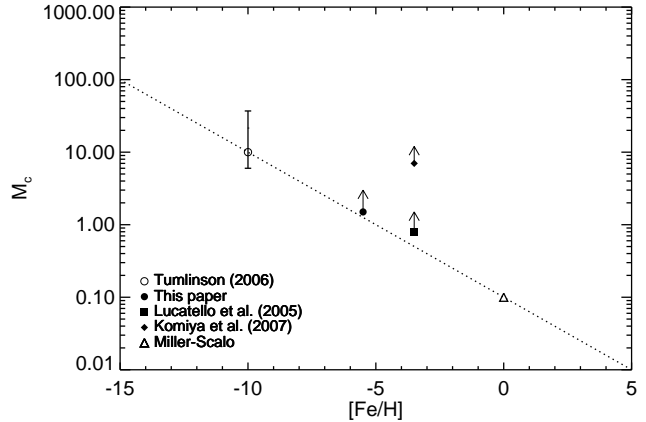}}}
\caption{The characteristic stellar mass $M_c$ for the log-normal IMF versus metallicity from four separate studies of IMF varying with metallicity. \label{IMF-mc-fig}} \end{figure}

The best explanation for the HMPs appears to be that they formed as the low-mass partners of a binary system drawn from a skewed IMF. This result corroborates the emerging evidence that the IMF at zero and low metallicity was skewed toward massive stars, which is a strong prediction of theoretical calculations of primordial star formation (Abel et al. 2002; Bromm et al. 1999, 2002; Yoshida et al. 2006; O'Shea et al. 2006). Tumlinson (2006) presented observational evidence of a top-heavy IMF for metal-free stars, which was based in part on star counts and chemical evolution constraints. That paper found a best-fit characteristic mass $M_c \simeq 10$ \msun\ but could not formally distinguish models with $M_c = 10 - 40$ \msun. The test case IMFs presented there can all reproduce the HMPs if stars up to 8 \msun\ can lead to C-rich HMPs (Figure 3).

Lucatello et al. (2005) also argued that an IMF skewed toward intermediate mass stars could explain the increase in the fraction of stars at low metallicity that are C-enriched. They did not specifically address origin scenarios for the HMPs, but the line of reasoning presented here for the implications of binarity in the HMPs for their IMF is closely related. They found that an IMF with $M_c = 0.79$ and $\sigma = 1.18$ explains the slightly increased incidence of C-rich stars among stars of \feh\ $= -2.0$ to $-3.0$. If one adopts this IMF for \feh\ $\sim -5.5$ as well, the probability of drawing the HMP stars is $\lesssim 1$\%, suggesting that the characteristic mass of the IMF continues to increase below \feh\ $\sim -3$. These results are summarized in Figure~\ref{IMF-mc-fig}, which shows the IMF results of T06, Lucatello et al. (2005), and the present study, together with the present-day Miller-Scalo IMF in the solar neighborhood ($M_c = 0.1$, $\sigma = 1.57$). Komiya et al. (2007) recently reported a new model of that attempts to match the observed frequency of carbon-rich EMP stars (Lucatello et al. 2006) with the same fundamental mechanism that Suda et al. (2004) invoked to explain HE0107-5240 and which is adopted here in modified form. Komiya et al. (2007) accurately reproduce the relative frequencies of carbon-enhanced and carbon normal EMP stars, with and without s-process enrichment, by invoking an IMF peaked at or above 6 \msun\ up to roughly [Fe/H] $\sim -2.5$. Their result is plotted as a lower limit in Figure~\ref{IMF-mc-fig}. The reader is referred to their paper for a thorough discussion of the underlying nucleosynthesis calculations and more insight into the relationship between the IMF and the CEMP phenomenon. Clearly more effort is needed to confirm these models and integrate them into our general understanding of chemical evolution and IMF formation at low metallicity.

Based on a critique of the proposed models for the two hyper-metal-poor stars, HE1327-2326 and HE0107-5240, I draw the following conclusions.
\begin{itemize}
\item[1.] The general category of models for HMPs that require dilution of the ejecta of one or a few supernovae into a large mass of primordial gas cannot reproduce the observed strong depletion of lithium in HE1327-2326. The faint supernova and massive star winds hypotheses appear to be excluded on this basis.
\item[2.] Of the three leading hypotheses that have been proposed for the origin of the two known hyper-metal-poor stars, HE0107-5240 and HE1327-2326, the one that best fits the overall dataset is the binary mass transfer hypothesis (Suda et al. 2004), in which the HMPs were formed as the low-mass partner in binary systems. This model matches all the observed abundance of the two HMPs, particularly the strong depletion of Li in HE1327-2326. The only exceptions are Mg and Al, for which nucleosynthesis in low-metallicity AGB stars is poorly understood.
\item[3.] If the binary hypothesis is confirmed by discovery of radial velocity variations in the two HMPs, their existence proves that low-mass stars were able to form at $Z \simeq 10^{-5.5} Z_{\odot}$. However, the statistically significant absence of stars with \feh\ $\sim -5$ and normal [C/Fe] implies that the stellar IMF at this metallicity was biased toward stars intermediate mass stars, with characteristic mass $M_c \gtrsim 1.4$ \msun\ required at $3\sigma$. Top-heavy IMFs with characteristic mass $\gtrsim 10$ \msun, such as those inferred for primordial stars (T06), also explain these results.
\item[4.] Taken together with the results of other studies of the stellar IMF at low metallicity, the HMP stars imply a steady increase of characteristic stellar mass on metallicity, with many other effects on the evolution of early galaxies.
\end{itemize}

To summarize, there is highly suggestive evidence that stars with low metallicity, \feh\ $\lesssim -2$, form in a mass function that is richer in intermediate mass stars than in low-mass stars. This surprising result is the exclusive outcome of stellar abundance studies, e.g. of ``Galactic archaeology''. The stellar IMF at metallicities below \feh\ $\lesssim -1$ cannot be addressed directly in galaxies in the local Universe, and may never be accessible directly for galaxies in the true conditions that obtained in the early universe. For the foreseeable future, local indicators like those used here will be the primary means for testing theories of star formation in the early phases of galaxy formation at $z > 6$.

\acknowledgements
This work owes to the teams of observers responsible for the discovery and characterization of the two HMP stars a debt that goes beyond what mere citations to their papers can convey. I am also pleased to acknowledge again the precursor work of the various theoretical groups whose work is discussed here. I gratefully acknowledge the generous support of Gilbert and Jaylee Mead for their namesake fellowship in the Yale Center for Astronomy and Astrophysics. Helpful comments by an anonymous referee provoked numerous clarifications and improvements.

\end{document}